\documentclass[aps,pra,superscriptaddress,showkeys,showpacs,nofootinbib,twocolumn]{revtex4-1}
\usepackage{amsmath}    
\usepackage{graphicx}   
\usepackage{array}
\usepackage[dvipdfx,cmyk,natbib]{xcolor}     
\usepackage[caption=false]{subfig}  
\usepackage[colorlinks,linkcolor=red,citecolor=brown,urlcolor=blue]{hyperref}
\usepackage{dsfont,bm}
\usepackage{amssymb}

\newcommand{\be}{\begin{equation}}
\newcommand{\ee}{\end{equation}}
\newcommand{\ben}{\begin{eqnarray}}
\newcommand{\een}{\end{eqnarray}}
\newcommand{\bes}{\begin{subequations}}
\newcommand{\ees}{\end{subequations}}
\newcommand{\bF}{\begin{figure}}
\newcommand{\eF}{\end{figure}}

\newcommand{\ttt}[1]{\text{#1}}
\newcommand{\bra}[1]{\langle{#1}\vert}
\newcommand{\ket}[1]{\vert{#1}\rangle}
\newcommand{\proj}[1]{\mbox{$|#1\rangle \!\langle #1 |$}}
\newcommand{\avg}[1]{\langle {#1} \rangle}

\begin{document}

\title{A quantum-inspired algorithm for estimating the permanent \\ of positive semidefinite matrices}

\author{L. Chakhmakhchyan}

\author{N. J. Cerf}

\author{R. Garcia-Patron}
\affiliation{Centre for Quantum Information and Communication, Ecole polytechnique de Bruxelles,
CP 165, Universit\'{e} libre de Bruxelles, 1050 Brussels, Belgium}

\date{\today}
\begin{abstract}
We construct a quantum-inspired classical algorithm for computing the permanent of Hermitian positive semidefinite matrices,
by exploiting a connection between these mathematical structures and the boson sampling model.
Specifically, the permanent of a Hermitian positive semidefinite matrix can be expressed in terms of the expected value of a random variable,
which stands for a specific photon-counting probability when measuring a linear-optically evolved random multimode coherent state.
Our algorithm then approximates the matrix permanent from the corresponding sample mean and is shown to run in polynomial time for various sets of
Hermitian positive semidefinite matrices, achieving a precision that improves over known techniques. This work illustrates how quantum optics
may benefit algorithms development.
\end{abstract}

\maketitle

\section{Introduction}\label{intr}
Linear quantum optics, which deals with the scattering of photons in a linear interferometer, is a promising candidate for the implementation of universal quantum computing~\cite{review}. This capability, first proved by the seminal scheme proposed by Knill, Laflamme and Milburn, makes use of passive linear-optics components, single-photon sources and detectors, as well as adaptive measurements~\cite{klm}. Recent considerable advances in the fabrication of reconfigurable optical circuits, single-photon sources and measurement devices of increasing reliability~\cite{sources, circuit} will make possible the design of fully integrated quantum optical circuits of tens to hundreds of qubits in the near future. However, due to the requirement of measurement-induced circuit control and the need to bring in ancillary photonic modes, among others, the current proof-of-principle realizations of this universal optical scheme are still far from a level where quantum advantage could be demonstrated. Surprisingly, as shown by Aaronson and Arkhipov, these highly demanding requirements can be relaxed within the boson sampling paradigm, while the resulting problem remains intractable for a classical computer~\cite{boson}. Namely, boson sampling is a task that consists of sampling from the probability distribution of detecting identical single photons at the output of a linear optical circuit. Despite the seeming simplicity of this task, the photon-counting probabilities are proportional to the squared modulus  of permanents of complex matrices~\cite{perm}, whose exact computation -- and even approximate estimation -- is in general an intractable ($\#$P-hard) task~\cite{scott, valiant}. This observation, along with several plausible complexity assumptions, is at the heart of the hardness proof of boson sampling. For this reason, boson sampling is now viewed as a very promising model to establish the advantage of a quantum computer over its classical counterpart, which has motivated a series of proof-of-principle experimental works~\cite{bs}.

Another striking feature of linear quantum optics, which we demonstrate in this paper, is its ability to inspire the construction of {\it efficient classical} algorithms. Specifically, building on the model of boson sampling with thermal states~\cite{ralph}, we propose an algorithm for estimating the permanent of Hermitian positive semidefinite matrices (HPSMs).
%
%
The algorithm exploits the optical equivalence theorem for multimode thermal states at the input of a linear optical circuit. Thermal states can be represented as a geometric distribution over Fock states, connecting the permanent of HPSMs with the single-photon measurement probabilities on the given multimode thermal state, evolved through an optical circuit~\cite{ralph}. At the same time, thermal states are also represented as Gaussian mixtures of coherent states~\cite{equiv}, an observation that allows us to construct an algorithm for approximating the permanent of HPSMs. Namely, up to a constant factor, our algorithm outputs the average (over a sample of polynomial size) of the probability of detecting the same single-photon pattern upon sending a Gaussian-distributed coherent state in a linear optical circuit.

The problem of approximating the permanent of HPSMs to within a multiplicative error has recently attracted the attention of computer scientists~\cite{scottblog}. In this regard, nothing is known as of today about the existence of polynomial-time techniques for approximating permanents of HPSMs with bounded relative errors~\cite{scottblog}, apart from that this can be achieved within the third level of the polynomial hierarchy (thus implying that the problem is not $\#$P-hard, unless the polynomial hierarchy collapses to its third level or beyond)~\cite{scottblog,ralph}. Further, we believe that, with the exception of  Gurvits' seminal algorithm~\cite{gurvits}, which approximates the permanent of any given $M\times M$ matrix with an additive error, no algorithm especially tailored to approximate the permanent of HPSMs has previously been developed. In this paper, we present such an algorithm, which substantially improves over Gurvits' technique in terms of additive errors and even achieves ``almost relative" error (i.e., proportional to the square root of the permanent itself) for some restricted set of HPSMs. Although our scheme does not resolve the open complexity-theoretic problem of approximating the permanent of HPSMs to within a multiplicative error~\cite{scottblog},  it further illustrates the potential of intertwining quantum optics with computer science. On a different note, let us point out that an algorithm for  estimating permanents of HPSMs may find applications in the assessment of boson sampling itself, in terms of generalized bunching~\cite{sch}. Namely, the latter allows one to efficiently certify that a physical device realizing boson sampling operates in the regime of full quantum coherence.

The rest of the paper is organized as follows. In the next section we recall the boson sampling model with thermal states and provide the basis for our algorithm. In Sec.~\ref{algo} we present the algorithm for approximating the permanent of Hermitian postive semidefinite matrices. Finally, in Sec.~\ref{concl} we draw our conclusions.

\section{Boson sampling with thermal states}\label{thermal}
As input state, we consider a multimode thermal state $\rho_{\mathrm{in}}^{\ttt{th}}= \bigotimes_{i=1}^M \rho_i^{\ttt{th}}$.
Each state $\rho_i^{\ttt{th}}$ is characterized in terms of its average photon number
$\avg{n_i}$ and can be expressed as an incoherent mixture of Fock states weighted by a geometric distribution with parameter
$\tau_i=\avg{n_i}/(\avg{n_i}+1)$, that is  $\rho_i^{\ttt{th}}=(1-\tau_i) \sum_{n=0}^\infty \tau_i^n \proj{n}$.

\begin{figure}[h!]
\begin{center}
\includegraphics[width=0.3\textwidth]{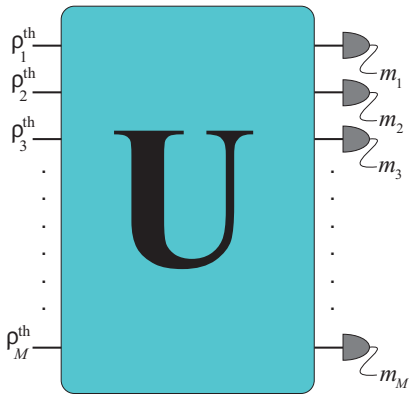}
\caption {Boson sampling setup with a $M$-mode thermal state $\rho_{\mathrm{in}}^{\ttt{th}}= \bigotimes_{i=1}^M \rho_i^{\ttt{th}}$
injected into a $M$-mode linear optical network that is characterized by the unitary matrix ${\bf U}$ (acting on bosonic mode operators),
followed by the detection of a pattern of photons $\{m_1, ..., m_M\}$. The problem is to sample from the probability distribution given by Eq.~(\ref{1}). \label{net}}
\end{center}
\end{figure}

The set of thermal states is injected into a  $M$-mode linear optical network
described by means of a $M\times M$ unitary matrix ${\bf U}$ that transforms the input mode operators $a_i^\dagger$
onto the output mode operators $b_i^\dagger$ (see also Fig.~\ref{net}):
\begin{equation}\label{2}
b_i^\dagger=\sum_{j=1}^M {\bf U}_{ij} \, a_j^\dagger.
\end{equation}
Exploiting the natural homomorphism between the $M\times M$ unitary matrix ${\bf U}$ and the corresponding unitary transformation $\mathcal{U}$
in state space~\cite{scott}, the linear optical evolution of $\rho_{\mathrm{in}}^{\ttt{th}}$ is given by $\rho_{\mathrm{out}}=\mathcal{U}\rho_{\mathrm{in}}^\ttt{th}\mathcal{U}^\dagger$,
and the joint probability of detecting $m_i$ photons on the $i$th output mode reads
\begin{equation}\label{1}
p^{\mathrm{th}}(\textbf{m})=\mathrm{Tr}[{\rho_{\mathrm{out}}| \textbf{m} \rangle \langle \textbf{m}|}],
\end{equation}
where $\ket{\textbf{m}}$ stands for a product of Fock states with $\textbf{m}\equiv \{m_1, ..., m_M\}$.
The boson sampling problem with thermal states consists of sampling from the probability distribution defined by Eq.~(\ref{1}),
given the set of input states $\rho_i^{\ttt{th}}$ and the transformation ${\bf U}$~\cite{ralph}.

We now restrict our analysis to a single element of the probability distribution $p^{\mathrm{th}}(\textbf{m})$ in Eq.~(\ref{1}), corresponding to the detection of a single photon in each output mode (i.e., $m_i=1$, $\forall i$), which reads (see Ref.~\cite{ralph} and the Appendix):
\begin{equation}\label{3}
p^{\mathrm{th}}\equiv p^{\mathrm{th}}(1, ..., 1)=\frac{ \mathrm{Per} \, {\bf A} }{\prod_{i=1}^M(1+\avg{n_i})}   ,
\end{equation}
where
\begin{eqnarray}\label{4.1}
&{\bf A=U D U^\dagger},& \\ \label{4.2}
&{\bf D}=\ttt{diag}\left\{\tau_1, ..., \tau_M \right\}&,
\end{eqnarray}
and the eigenvalues satisfy
\begin{equation}
1>\tau_i=\avg{n_i}/(\avg{n_i}+1)\geq 0, \,\,\, \forall i. \label{4.3}
\end{equation}
This connection between the probability of detecting a pattern of $M$ single photons at the output of the linear-optical evolution of a $M$-mode thermal state $\rho_{\mathrm{in}}^{\ttt{th}}$ and the permanent of a $M\times M$ Hermitian positive semidefinite matrix ${\bf A}$ of bounded eigenvalues ($\tau_i<1$), is one of the two main ingredients in the construction of our algorithm. It is also worth pointing that the unitary that diagonalizes ${\bf A}$ is precisely the one that describes the circuit itself, while the spectrum of ${\bf A}$ is determined in terms of the average photon numbers $\avg{n_i}$ of the input thermal states.

\section{Algorithm for estimating permanents of Hermitian positive semidefinite matrices}\label{algo}

\subsection{The main intuition behind our algorithm}
The next key ingredient in the construction of our algorithm is to exploit the Glauber-Sudarshan $P$-representation~\cite{equiv} to write down the $M$-mode input thermal state $\rho_{\text{in}}^\text{th}$ as a mixture of $M$-mode coherent states $\ket{\pmb{\alpha}}\equiv\ket{\alpha_1, ..., \alpha_M}=\bigotimes_{i=1}^M \ket{\alpha_i}$ according to a Gaussian distribution
\begin{equation}
\rho_{\text{in}}^\text{th}=\int_{\mathbb{C}^M} \prod_{i=1}^M \left[\frac{d^2\alpha_i }{\pi\avg{n_i}}  \, \exp\left(-\frac{|\alpha_i|^2}{\avg{n_i}}\right)\right] \proj{\pmb{\alpha}}.
\end{equation}
Consequently, one can express the linear optical evolution of the input state in terms of that of their component coherent states $\ket{\pmb{\alpha}}$. Namely, $\mathcal{U}$ transforms a tensor product of coherent states $\bigotimes_{i=1}^M \ket{\alpha_i}$ into another tensor product of coherent states $\bigotimes_{i=1}^M \ket{\beta_i}$ with amplitudes
\begin{equation}\label{5}
\beta_i=\sum_{j=1}^M {\bf U}_{ji} \alpha_j.
\end{equation}
In other words, coherent states remain in a tensor product form while evolved through a linear optical circuit. Thus, the joint probability $p^{\mathrm{cs}}(\pmb{\alpha})\equiv p^{\mathrm{cs}}(\alpha_1,..., \alpha_M)$  of detecting a single photon at each output mode, with a $M$-mode coherent state $\ket{\pmb{\alpha}}$ at the input, admits a simple product form
\begin{equation}\label{6}
p^\ttt{cs}(\pmb{\alpha})=\prod_{i=1}^{M}e^{-|\beta_i|^{2}}|\beta_i|^{2},
\end{equation}
where the dependence on $\alpha_i$'s is implicit via Eq.~(\ref{5}). As a consequence, the probability $p^{\mathrm{th}}$ of Eq.~(\ref{3}) is alternatively represented as
\ben\label{7}
p^\text{th}=\int_{\mathbb{C}^M} \prod_{i=1}^M \frac{d^2\alpha_i }{\pi\avg{n_i}} \exp\left(-\frac{|\alpha_i|^2}{\avg{n_i}}\right)  p^\ttt{cs}(\pmb{\alpha}). \,\, ~
\een
Therefore, we end up with the expected value of the function $p^\ttt{cs}(\pmb{\alpha})$ of random variables $\alpha_1,..., \alpha_M$, which results from integrating $p^\ttt{cs}(\pmb{\alpha})$ over a complex Gaussian distributed probability measure over the complex random variables $\alpha_i\in \mathcal{N}_\mathbb{C}(0, \avg{n_i})$ [or for the sake of shortness, $\pmb{\alpha}\in \mathcal{N}$, with $\mathcal{N}$ denoting the set $\{\alpha_i\in \mathcal{N}_\mathbb{C}(0, \avg{n_i}), i=1, ..., M\}$], i.e., $p^\text{th}=\mathbb{E}[p^\ttt{cs}(\pmb{\alpha})]_{\pmb{\alpha}\in \mathcal{N}}$. Finally, exploiting the connection between $p^\text{th}$ and $\mathrm{Per} \, {\bf A}$ in Eq.~(\ref{3}) we rewrite the permanent of any HPSM $\bf A$ satisfying Eqs.~(\ref{4.1})-(\ref{4.3}) as
\ben\label{8}
\mathrm{Per}{\bf A}&=&\frac{1}{\prod_{i=1}^M(1-\tau_i)}\mathbb{E}[p^\ttt{cs}(\pmb{\alpha})]_{\pmb{\alpha}\in \mathcal{N}}.
\een
A sampling algorithm approximating $\mathbb{E}[p^\ttt{cs}(\pmb{\alpha})]_{\pmb{\alpha}\in \mathcal{N}}$ will thus immediately translate into an algorithm which, according to the law of large numbers, converges to $\mathrm{Per} \, {\bf A}$.

\subsection{The algorithm}
In order to approximate the permanent of an arbitrary HPSM matrix ${\bf \Lambda}$ we first diagonalize ${\bf \Lambda}={\bf U} {\bf D} {\bf U^\dagger}$, where the unitary matrix ${\bf U}$ encodes the eigenvectors and ${\bf D}=\mathrm{diag}(\lambda_1, ..., \lambda_M)$, the spectrum of $\pmb{\Lambda}$. If the largest eigenvalue does not satisfy $\lambda_{\ttt{max}}<1$, we re-scale the matrix, ${\bf A}={\pmb \Lambda}/(C\lambda_\ttt{max})$, reducing the problem to finding the permanent of ${\bf A}$, knowing that $\ttt{Per} \, \pmb{\Lambda}=(C \lambda_\ttt{max})^M \, \ttt{Per {\bf A}}$. Remark that $C>1$ is a specific constant that is necessary to avoid the divergence of the highest $\avg{n_i}= \tau_i/(1-\tau_i)$, while it also provides a certain tunability for the algorithm. Consequently, we can write
$\mathrm{Per}\, {\pmb \Lambda}=Z\mathbb{E}[p^\ttt{cs}(\pmb{\alpha})]_{\pmb{\alpha}\in \mathcal{N}}$, with
\begin{eqnarray}\label{10}
Z=\frac{(C\lambda_\mathrm{max})^{2M}}{\prod_{i=1}^M
\left(C\lambda_{\text{max}}-\lambda_i\right)},
\end{eqnarray}
where we expressed the parameters $\tau_i$ in terms of the eigenvalues of $\pmb{\Lambda}$. The second step of the algorithm then consists in approximating $\mathbb{E}[p^\ttt{cs}(\pmb{\alpha})]_{\pmb{\alpha}\in \mathcal{N}}$ as follows:
\begin{itemize}
\item  For the selected  error $\epsilon$ in the approximation of $\mathrm{Per}\, {\pmb \Lambda}$
and for a probability $\delta$ of failure of the algorithm, calculate the needed number of samples $N$, using Eq.~(\ref{eq:samples}) below.
\item Generate $N$ samples $\pmb{\alpha}^{(j)}$ ($j=1,..., N$) of the random string $\pmb{\alpha}=\{\alpha_1, ..., \alpha_M\}$, where each $\alpha^{(j)}_i$ is drawn at random from the Gaussian distribution $\mathcal{N}_\mathbb{C}(0, \avg{n_i})$.
\item For each string $\pmb{\alpha}^{(j)}$, by means of Eqs.~(\ref{5}) and (\ref{6}), calculate $p^{\mathrm{cs}}[\pmb{\alpha}^{(j)}]$.
\item Calculate the sample mean
\begin{equation}\label{9}
\mu=\sum_{j=1}^Np^{\mathrm{cs}}[\pmb{\alpha}^{(j)}]/N.
\end{equation}
\item Finally, output $Z\cdot \mu$
\end{itemize}

Our algorithm involves several computational steps whose running time is polynomial in $M$. First, computing the eigenvalues and eigenvectors of the $M\times M$ HPSM can be done in time $O(M^3)$ with the traditional QR or divide-and-conquer algorithms \cite{eigen}. Next, for each sample of $p^{\mathrm{cs}}(\pmb{\alpha})$, one
has to generate a random string $\pmb{\alpha}=\{\alpha_1, ..., \alpha_M\}$ with $\alpha_i \in \mathcal{N}_\mathbb{C}(0, \avg{n_i})$, for which efficient sampling techniques from Gaussian distributions are available \cite{sample}. Afterwards,
one multiplies the matrix ${\bf U}$ with the column of $\alpha_i$'s yielding the amplitudes $\beta_i$'s in time
$O(M^2)$.  If one generates $N$ samples of $p^\mathrm{cs}(\pmb{\alpha})$, then the overall running time of the algorithm scales as $O(M^2[M+N])$.

\subsection{Error analysis and scaling of the algorithm}
In order to determine the efficiency of our algorithm, as well as the scaling of its running time with respect to the matrix size $M$, we have to estimate the number of samples $N$, needed to reach a given precision in the approximation of $\ttt{Per}\pmb{\Lambda}=Z\mathbb{E}[p^\ttt{cs}(\pmb{\alpha})]_{\pmb{\alpha}\in \mathcal{N}}$.
Remark that as far as relative-error analysis is concerned, the prefactor $Z$ in front of $\mathbb{E}[p^\ttt{cs}(\pmb{\alpha})]_{\pmb{\alpha}\in \mathcal{N}}$ [Eq.~(\ref{10})] is irrelevant. In contrast, for the additive-error approximation, it should be carefully taken into account (see the Appendix).

In order to estimate the sample size $N$ needed, we make use of the Hoeffding inequality~\cite{chernoff, chernoff1}. The latter provides an upper bound for the probability of the sample mean $\mu$ to deviate from the expected value
$\mathbb{E}[p^\ttt{cs}(\pmb{\alpha})]_{\pmb{\alpha}\in \mathcal{N}}$, given the sample size $N$ and the constraint $0 \leq p^\ttt{cs}(\pmb{\alpha})\leq e^{-M}$. Translated into the estimate $Z\cdot \mu$ of Per$\pmb{\Lambda}$, Hoeffding inequality yields:
\begin{equation}\label{27}
\mathrm{Pr}(|\mathrm{Per}\,{\pmb\Lambda}-Z\mu| \geq Z\epsilon)\leq \exp\left(-\frac{2 N \epsilon^2}{e^{-2M}}\right).
\end{equation}

Denoting by $\delta$ the failure probability of the algorithm, we find the sample size $N$ that results in an error $\epsilon$ for $\mathrm{Per}\,{\pmb\Lambda}$,
\begin{equation}\label{eq:samples}
N=\frac{Z^2e^{-2M}}{2\epsilon^2}\ln\frac{1}{\delta}.
\end{equation}
For the algorithm running time to scale polynomially, the sample size $N$ should stay polynomial in the matrix size $M$, imposing conditions
on the spectra of the HPSM, as we detail below.

Up to our knowledge the only algorithm capable of approximating the permanent of HPSMs is Gurvits' algorithm, which is defined for general complex matrices. Gurvits' algorithm exploits the fact that the permanent of any matrix {\bf X} can be written down as the expected value of efficiently computable bounded random variables~\cite{ryser}. This enables approximating Per{\bf X} in terms of the corresponding sample mean in time $O(M^2/\varepsilon^2)$, yielding an additive error $\pm \varepsilon ||{\bf X}||^M$, where $||{\bf X}||$ denotes the trace norm of {\bf X}, which reads $\pm \varepsilon \lambda_\text{max}^M$ for HPSMs. As we detail in the Appendix, analyzing the additive error of our scheme we find a set ($S1$) of cases where our technique outperforms Gurvits' algorithm for HPSMs. Namely, we find a specific regime where the additive error of our algorithm decreases exponentially faster than that of Gurvits, at the price of a small polynomial overhead.
More precisely, we achieve the error $\pm \varepsilon l^M \lambda_\ttt{max}^M$ ($l\leq1$) for the set of matrices such that their spectra satisfy the following necessary and sufficient condition
\begin{equation}
\sqrt[M]{\prod_{i=1}^M \left(1-\frac{\lambda_i}{C\lambda_{\text{max}}}\right)}\geq\frac{C}{e}.
\end{equation}

For yet another set ($S2$) of HPSMs, satisfying similar constraints but also the condition $\lambda_{{\rm max}}>1$, our scheme yields an additive error decreasing exponentially with $M$, where the Gurvits' algorithm fails to do so.
Interestingly, as we detail in the Appendix, our derivation yields as a corollary an upper-bound for the permanent of HPSMs in $S2$. It implies an exponential decrease of the permanent with $M$, where Glynn's formula fails to do so for matrices satisfying $\lambda_{{\rm max}}>1$, as Glynn's formula leads to the upper bound $\text{Per}\pmb{\Lambda}\leq \lambda_\text{max}^M$~\cite{gurvits1}.

Finally, we are also able to achieve an ``almost-relative" error $\pm \varepsilon \sqrt{\text{Per}\pmb{\Lambda}}$ for a different restricted class ($S3$) of HPSMs. The corresponding condition relies again on the spectral properties of the matrix $\pmb{\Lambda}$.

\section{Conclusion}\label{concl}

We have presented a quantum-inspired algorithm, which exploits tools from quantum optics
to address a classical computational problem -- estimating the permanent of Hermitian positive semidefinite matrices. By use of Monte-Carlo type technique, the permanent is approximated as the expected value of a random variable, up to a prefactor that only depends on the spectrum of the matrix. Interestingly, this random variable finds a natural physical interpretation as it stands for the joint probability of detecting a single photon at the output of a specific linear-optical circuit, injected with a $M$-mode coherent state of normally-distributed random amplitudes. Additionally, the unitary defining the circuit is the one that diagonalizes the given Hermitian positive semidefinite matrix, and the eigenvalues are connected to the variance of the normal distribution of the $M$-mode coherent state.

The error analysis shows, for a specific set of Hermitian positive semidefinite matrices, that our polynomial-time algorithm yields better additive errors than Gurvits' technique. Moreover, for a restricted class of Hermitian positive semidefinite matrices, we are even able to achieve an ``almost-relative'' error, proportional to the square root of the permanent itself. We believe that the necessary conditions developed in the appendix indicate that
these restricted sets of matrices do not reduce to computationally trivial classes (with respect to the permanent computation), 
but a full analysis should be carried out in order to confirm it. Whether these restrictions should be viewed as a caveat of the proposed algorithm is left for future work, but we stress that this is, up to our knowledge, the first classical algorithm especially tailored to approximate the permanent of Hermitian positive semidefinite matrices.

We hope that this work will motivate further investigation to develop a multiplicative-error approximation algorithm of the permanent of Hermitian positive semidefinite matrices, a complexity-theoretic question that remains open. We also believe that our work highlights the benefits that exploiting the connection between the theory of computer science and quantum optics could bring to both communities.

\section*{Acknowledgments} We thank Anthony Leverrier for useful discussions and comments. This work was supported by H2020-FETPROACT-2014 Grant QUCHIP (Quantum Simulation on a Photonic Chip; grant agreement no. 641039, http://www.quchip.eu). R.G.-P. acknowledges financial support as a research associate of the Fonds de la Recherche Scientifique (F.R.S.-FNRS, http://www.fnrs.be).


\appendix

\section{Efficient regimes and error analysis of the algorithm approximating the permanent of Hermitian positive semdefinite matrices}\label{errors}

In this section we detail the efficient regimes of the proposed algorithm and estimate its failure probability. As already mentioned in the main text, the running time of our scheme strongly depends on the  sample size $N$ which approximates the expected value $\mathbb{E}[p^\ttt{cs}(\pmb{\alpha})]_{\pmb{\alpha}\in \mathcal{N}}$ and thus the permanent of a given HPSM. And in order to estimate $N$ we make use of the Hoeffding inequality. It is applicable in our case, since the random variable $p^{\mathrm{cs}}(\pmb{\alpha})$ is bounded. Namely, due to its definition $p^{\mathrm{cs}}(\pmb{\alpha})\leq \prod_{i=1}^M e^{-1}=e^{-M}$, which means that the random variable $p^{\mathrm{cs}}(\pmb{\alpha})$ (and thus its expected value), lies within the interval $[0, e^{-M}]$. Thus, the Hoeffding inequality provides an upper bound for the probability of the approximant sample mean $\mu$ to be far from the expected value $\mathbb{E}[p^\ttt{cs}(\pmb{\alpha})]_{\pmb{\alpha}\in \mathcal{N}}$, given the sample $N$ and the fact that $0 \leq p^{\mathrm{cs}}(\pmb{\alpha})\leq e^{-M}$:
\begin{equation}\label{27}
\mathrm{Pr}(|\mathbb{E}[p^\ttt{cs}(\pmb{\alpha})]_{\pmb{\alpha}\in \mathcal{N}}-\mu| \geq \epsilon)\leq \exp\left(-\frac{2 N \epsilon^2}{e^{-2M}}\right).
\end{equation}
Therefore, given the failure probability of the algorithm $\delta$, from the above equation we find that the sample size $N$, which results in an error $\epsilon$ is:
\begin{equation}\label{28}
N=\frac{ e^{-2M}}{2\epsilon^2}\ln\frac{1}{\delta}.
\end{equation}
We also restate the relation between the expected value $\mathbb{E}[p^\ttt{cs}(\pmb{\alpha})]_{\pmb{\alpha}\in \mathcal{N}}$ and the permanent of any given HPSM $\pmb{\Lambda}$, in a form more suitable for the further analysis:
\be\label{28.1}
\text{Per}\pmb{\Lambda}=Z \mathbb{E}[p^\ttt{cs}(\pmb{\alpha})]_{\pmb{\alpha}\in \mathcal{N}}
\ee
with
\begin{eqnarray}\label{29}
&& Z=\frac{C^{2M} \lambda_\mathrm{max}^{2M}}{a^M}, \\ \label{29.11}
&& a=\sqrt[M]{\prod_{i=1}^M \left(C\lambda_{\text{max}}-\lambda_i\right)}.
\end{eqnarray}
In the above equation $a$ is the geometric mean of the quantities $\{C\lambda_{\text{max}}-\lambda_1,..., C\lambda_{\text{max}}-\lambda_M \}$, which, combined with the Inequality of arithmetic and geometric means, satisfies
\be\label{30}
\lambda_\text{max}(C-1)\leq a \leq C \lambda_\text{max}-\bar{\lambda},
\ee
where
\be
\bar{\lambda}=\frac{1}{M}\sum_{i=1}^M\lambda_i,
\ee
is the eigenvalue mean.

It is important to note here that since the permanent of the HPSM $\pmb{\Lambda}$ is equal the expected value of $p^\ttt{cs}(\pmb{\alpha})$ times the constant $Z$, the approximation of $\mathbb{E}[p^\ttt{cs}(\pmb{\alpha})]_{\pmb{\alpha}\in \mathcal{N}}$ to within an additive error $\epsilon$ results in an error $Z\epsilon$ for $\ttt{Per}\pmb{\Lambda}$. We proceed now with the analysis of several regimes of our algorithm approximating  $\ttt{Per}\pmb{\Lambda}$ and yielding distinct types of additive errors.

\subsection{Additive error beating Gurvits' algorithm (set $S1$ of the main text)}\label{add3}

Firstly, we compare the error results provided by our algorithm to that of the Gurvits' one. The latter, having running time $O(M^2/\varepsilon^2)$, estimates the permanent of any $M\times M$ matrix  ${\bf X}$, to within an additive error $\varepsilon ||{\bf X}||^M$ ($||{\bf X}||$ is the spectral norm of ${\bf X}$). As for HPSMs $||{\pmb \Lambda}||=\lambda_\mathrm{max}$, Gurvits' additive error reads $\varepsilon \lambda_\mathrm{max}^M$.

Consequently, the requirement that our algorithm results in an exponentially smaller error than that of Gurvits' (with a polynomial overhead), i.e., an additive error $\varepsilon (l \lambda_\mathrm{max})^M$ for the permanent approximation, imposes us to set $\epsilon=\varepsilon l^M \lambda_\text{max}^M/Z$ in Eq.~(\ref{28}):
\be\label{31.2}
N=\frac{1}{2\varepsilon^2}\ln\frac{1}{\delta} \left(\frac{\lambda_{\text{max}} C^2}{l e a}\right)^{2M},
\ee
together with the additional constraint $l\leq 1$. Now, our aim is to approximate the permanent of the matrix $\pmb{\Lambda}$ to within a minimal additive error within this regime. Therefore, in order to avoid the exponential increase of the sample size $N$ we impose the following conditions:
\begin{equation}\label{34}
l\leq 1, \,\,\,\,\, \frac{\lambda_{\text{max}} C^2}{e a}\leq l,
\end{equation}
leading to the inequality
\begin{equation}\label{ap3}
a\geq\frac{\lambda_{\text{max}} C^2}{e}.
\end{equation}
This inequality defines a condition on the spectra of the HPSMs that is necessary and sufficient to guarantee the efficiency of our approximation algorithm [for achieving the additive error $\varepsilon (l \lambda_\mathrm{max})^M$, exponentially smaller in $M$ than that of Gurvits' algorithm]. It is easy to see that conditions (\ref{34}) can be recovered from (\ref{ap3}) by setting $l=\lambda_{\text{max}} C^2/(e a)$.

Meanwhile, combinig the upper bound of $a$ in Eq.~(\ref{30}) with  Eq.~(\ref{ap3}), we find the
necessary condition
\begin{equation}\label{ap4}
\bar{\lambda}\leq \lambda_{\text{max}} C \left(1-\frac{C}{e}\right)\leq\frac{e}{4}\approx 0.680,
\end{equation}
which provides some intuition on the regime of parameters where our algorithm improves over Gurvits' algorithm.
Furthermore, since $\bar{\lambda}\geq0$, we obtain the second necessary condition
\begin{equation}\label{ap5}
C\leq e.
\end{equation}
It is also possible to bound $l$ from below:
\be\label{35.3}
l \geq\frac{C^2\lambda_\text{max}}{e(C \lambda_\text{max}-\bar{\lambda})}\geq \frac{1}{e},
\ee
giving information on the minimal additive error that our scheme potentially provides within the regime discussed in this subsection.

Finally, it is worth noting that if the maximal eigenvalue of the given HPSM $\pmb{\Lambda}$ is smaller than one, $\lambda_\ttt{max}<1$, one does not require the rescaling of $\pmb{\Lambda}$.  In other words, there is no necessity of dividing it by $C \lambda_\mathrm{max}$. This effectively corresponds to setting $C\lambda_\text{max}=1$ (or replacing $C$ by $1/\lambda_\ttt{max}$). Therefore, the conditions (\ref{ap3})-(\ref{ap5}) can be readily applied for HPSMs with $\lambda_\ttt{max}<1$, by simply replacing $C$ by $1/\lambda_\ttt{max}$ [and doing that in the definition (\ref{29.11}) as well]. Remark that when $\lambda_\ttt{max}<1$, the additive error $\varepsilon \lambda_\ttt{max}^M$ of the Gurvits' algorithm itself is exponentially decreasing in $M$.

\subsection{Exponentially decreasing additive error: $\lambda_\text{max}\geq1$ (set $S2$ of the main text)}\label{add1}

In this subsection we show that our algorithm is capable of providing additive error results well beyond that of the Gurvits' scheme. Namely, we aim at achieving an additive error that decreases exponentially in $M$ for HPSMs with $\lambda_\ttt{max} \geq 1$, while Gurvits' error, $\varepsilon \lambda_\ttt{max}$, is exponentially increasing in this case. In other words, we wish to guarantee an additive error $\varepsilon k^M$ ($k\leq 1$) for the permanent of the HPSM $\pmb{\Lambda}$. Therefore, in Eq.~(\ref{28}) we set $\epsilon=\varepsilon k^M/A$:
\be\label{31}
N=\frac{1}{2\varepsilon^2}\ln\frac{1}{\delta} \left(\frac{\lambda_{\text{max}}^2 C^2}{k e a}\right)^{2M}.
\ee
Our aim is then to approximate the permanent of the matrix $\pmb{\Lambda}$ to within the minimal additive error, which decreases exponentially in $M$. Thus, in order to avoid the exponential increase of the sample size $N$, analogous to the prevous subsection, we obtain
\begin{equation}\label{ap0.1}
a\geq\frac{\lambda_{\text{max}}^2 C^2}{e}.
\end{equation}
This inequality defines a necessary and sufficient condition on the spectra of a given HPSM, which guarantees the efficiency of our approximation algorithm within the present regime.

On the other hand, due to Eq.~(\ref{30}), $a\leq C \lambda_\text{max}-\bar{\lambda}$ and thus along with Eq.~(\ref{ap0.1}) we find the necessary condition
\begin{equation}\label{ap0.2}
\bar{\lambda}\leq \lambda_{\text{max}} C \left(1-\frac{\lambda_{\text{max}}C}{e}\right).
\end{equation}
Meanwhile, since $\bar{\lambda}\geq0$, we end up with another necessery condition,
\begin{equation}\label{ap0.3}
\lambda_{\text{max}}\leq\frac{e}{C}.
\end{equation}
The constraints (\ref{ap0.2}) and (\ref{ap0.3}) thus provide some intuition on the spectrum of HPSMs for which our algorithm yields an exponentially decreasing error in $M$ (in polynomial time). And as $\lambda_\text{max}\geq1$, we obtain the next necessary condition,
\begin{equation}\label{ap0.333}
C\leq e.
\end{equation}
In other words, for the permanent of HPSMs with $1\leq\lambda_\text{max}<e$, we are potentially able to attain an exponentially decreasing error $\varepsilon k^M$.

Finally, as in the previous subsection, we are  able to bound $k$ from below:
\be
k\geq\frac{C^2\lambda_\text{max}^2}{e(C \lambda_\text{max}-\bar{\lambda})}\geq \frac{1}{e},
\ee
which gives information on the minimal additive error that our scheme potentially provides within the regime discussed here.

\subsection{``Almost relative" error (set $S3$ of the main text)}\label{errors.1}

In this subsection we analyze the efficiency of our scheme beyond approximations to within an additive error. Namely, we consider the task of estimating the permanent of an HPSM $\pmb{\Lambda}$ to within an error $\pm \varepsilon \sqrt{\text{Per}\pmb{\Lambda}}$. For that, in Eq.~(\ref{28}) we set $\epsilon=\varepsilon\sqrt{\mathbb{E}[p^\ttt{cs}(\pmb{\alpha})]_{\pmb{\alpha}\in \mathcal{N}}/Z}$, yielding:
\be\label{35.13}
N=\frac{1}{2\varepsilon^2}\ln\frac{1}{\delta} \frac{Z}{e^{2M}\mathbb{E}[p^\ttt{cs}(\pmb{\alpha})]_{\pmb{\alpha}\in \mathcal{N}}}.
\ee
In order to reveal the efficient regimes of this specific case, we firstly provide a lower bound for $\mathbb{E}[p^\ttt{cs}(\pmb{\alpha})]_{\pmb{\alpha}\in \mathcal{N}}$. Thus, we return to the definition of $p^\ttt{cs}(\pmb{\alpha})$ itself:

\begin{widetext}
\begin{eqnarray}\label{36}\nonumber
\mathbb{E}[p^\ttt{cs}(\pmb{\alpha})]_{\pmb{\alpha}\in \mathcal{N}}&=&\int_{\mathbb{C}^M}d \pmb{\alpha}  \prod_{j=1}^M e^{-|\beta_j|^{2}} |\beta_j|^{2} \prod_{i=1}^M\left[\frac{1}{\pi\avg{n_i}}\exp\left(-\frac{|\alpha_i|^2}{\avg{n_i}}\right)\right]=
\frac{1}{\pi^M\prod_{i=1}^M\avg{n_i}}\int_{\mathbb{C}^M}d \pmb{\alpha} \left[\prod_{j=1}^M e^{-|\beta_j|^{2}} |\beta_j|^{2}\right. \times \\
&& \left.\exp\left(-\sum_{i=1}^M\frac{|\alpha_i|^{2}}{\avg{n_i}}\right)\right] \geq\frac{1}{\pi^M\prod_{i=1}^M\avg{n_i}}\int_{\mathbb{C}^M}d \pmb{\alpha} \prod_{j=1}^M e^{-|\beta_j|^{2}} |\beta_j|^{2} \exp\left(-\frac{1}{\avg{n_\mathrm{min}}}\sum_{i=1}^M|\alpha_i|^{2}\right),
\end{eqnarray}
\end{widetext}
where $d\pmb{\alpha}\equiv d^2\alpha_1...d^2\alpha_M$ and $\avg{n_\mathrm{min}}$ denotes the minimal $\avg{n_i}$:
\begin{eqnarray}\label{37}
\avg{n_\mathrm{min}}=\frac{\lambda_\mathrm{min}}{C \lambda_\mathrm{max} -\lambda_\mathrm{min}}.
\end{eqnarray}
Next, taking into account that $\beta_i=\sum_{j=1}^M {\bf U}_{ji} \alpha_j$ and the fact that the matrix ${\bf U}$ is unitary, we find that $\sum_{i=1}^M|\alpha_i|^{2}=\sum_{j=1}^M|\beta_j|^{2}$. Thus Eq.~(\ref{36}) reads
\begin{widetext}
\begin{eqnarray}\label{36.1}
\mathbb{E}[p^\ttt{cs}(\pmb{\alpha})]_{\pmb{\alpha}\in \mathcal{N}}\geq \frac{1}{\pi^M\prod_{i=1}^M\avg{n_i}}\int_{\mathbb{C}^M}d \pmb{\beta} \prod_{i=1}^M e^{-|\beta_i|^{2}} |\beta_i|^{2} \exp\left(-\frac{1}{\avg{n_\mathrm{min}}}\sum_{i=1}^M|\beta_i|^{2}\right)=\frac{1}{\prod_{i=1}^M\avg{n_i}}\left(\frac{\avg{n_\mathrm{min}}}{1+\avg{n_\mathrm{min}}}\right)^{2 M}.\,\,\,\,\,
\end{eqnarray}
\end{widetext}
In the above equation we also made a change of integration variables (from $\{\alpha_1, ..., \alpha_M\}$ to $\{\beta_1, ..., \beta_M\}$), using the fact that the absolute value of the Jacobian determinant of the corresponding unitary transformation is unity ($|\mathrm{det}{\bf U}|=1$) \cite{kaplan}. Finally, we rewrite the above lower bound in terms of the eigenvalues of $\pmb{\Lambda}$, yielding
\begin{eqnarray}\label{39}
\mathbb{E}[p^\ttt{cs}(\pmb{\alpha})]_{\pmb{\alpha}\in \mathcal{N}}\geq&&\frac{1}{\prod_{i=1}^M\avg{n_i}}\left(\frac{\lambda_\text{min}}{C \lambda_\text{max}}\right)^{2M}=\\ \nonumber &&\prod_{i=1}^M \frac{C \lambda_\mathrm{max} -\lambda_i}{\lambda_i} \left(\frac{\lambda_\text{min}}{C \lambda_\text{max}}\right)^{2M}.
\end{eqnarray}
Substituting this expression into Eq.~(\ref{35.13}) we find
\begin{widetext}
\ben\label{35.14}
N\leq\frac{1}{2\varepsilon^2}\ln\frac{1}{\delta}\cdot \frac{1}{e^{2M}}\frac{\lambda_\mathrm{max}^{2M} C^{2M}}{\prod_{i=1}^M(C\lambda_\text{max}-\lambda_i)} \frac{\prod_{i=1}^M \lambda_i}{\prod_{i=1}^M(C \lambda_\mathrm{max} -\lambda_i)}\left(\frac{C \lambda_\text{max}}{\lambda_\text{min}}\right)^{2M}=\left(\frac{\lambda_{\text{max}}^4 C^4 d}{\lambda_{\text{min}}^2 e^2}\right)^{M},
\een
\end{widetext}
where
\be\label{35.15}
d=\sqrt[M]{\prod_{i=1}^M \frac{\lambda_i}{(C\lambda_\text{max}-\lambda_i)^2}}
\ee
is the geometric mean of the quantities $\{\lambda_1/(C\lambda_\text{max}-\lambda_1)^2, ..., \lambda_M/(C\lambda_\text{max}-\lambda_M)^2\}$. As in the previous cases, the algorithm is efficient if
\be \label{35.16}
\frac{\lambda_{\text{max}}^4 C^4 d}{\lambda_{\text{min}}^2 e^2}\leq1.
\ee
The set of matrices defined by the above inequality is not empty, i.e. there exist HPSMs such that their spectrum satisfies the condition (\ref{35.16}), which, in turn, guarantees the additive error $\pm \varepsilon \sqrt{\ttt{Per}\pmb{\Lambda}}$ of our algorithm. This condition has to be checked for a given matrix $\pmb{\Lambda}$.

As for the special case $\lambda_\text{max}<1$, as already mentioned above, we readily obtain the corresponding constraints by simply replacing $C$ by $1/\lambda_\ttt{max}$:
\be \label{35.17}
\frac{f}{\lambda_{\text{min}}^2 e^2}\leq1,
\ee
where
\be\label{35.18}
f=\sqrt[M]{\prod_{i=1}^M \frac{\lambda_i}{(1-\lambda_i)^2}}.
\ee

As a conclusion to this section we note that depending on the type of the error one wants to achieve, the corresponding conditions (\ref{ap3}); (\ref{ap0.1}); (\ref{35.16}) [or (\ref{35.17})] should be checked, before applying the steps of the algorithm outlined in the main text of the paper. If satisfied, one proceeds with the estimation scheme. Finally, we also emphasize that our method possesses certain tunability in terms of the parameter $C$, which appears in the corresponding conditions. It allows one to expand the applicability of the algorithm, as well as to the optimize the resulting error. Nevertheless, as our analysis shows, $C$ cannot be chosen arbitrarily large. Namely, in order to avoid the exponential increase of the corresponding additive error, one has to set $C\leq e$.

\section{Permanent upper and lower bounds}

It is worth noting an important corollary of our results, which provides an exponentially decreasing upper bound for the permanent of HPSMs. Namely, combining Eqs.~(\ref{28.1})-(\ref{29}) with the fact that $p^{\mathrm{cs}}(\pmb{\alpha})\leq e^{-M}$, we find:
\begin{eqnarray}\label{43}
\mathrm{Per}\pmb{\Lambda} \leq \left(\frac{C^2 \lambda_\mathrm{max}^2}{a e}\right)^{M}.
\end{eqnarray}
Therefore, if the matrix $\pmb{\Lambda}$ satisfies the conditions $\lambda_\ttt{max}\geq 1$ and (\ref{ap0.1}), $\mathrm{Per}\pmb{\Lambda}$ decreases exponentially with the size of the matrix $M$. In contrast, the standard upper bound for permanents, resulting from Glynn's formula, yields $\text{Per}\pmb{\Lambda}\leq \lambda_\text{max}^M$, which {\it increases} exponentially in $M$ if $\lambda_\ttt{max}\geq 1$.

Additionally, we are also able to provide a novel lower bound for the permanent of HPSMs, expressed in terms their spectra, using the lower bound for $p^{\mathrm{cs}}(\pmb{\alpha})$ of Eq.~(\ref{39}):

\begin{eqnarray}\label{43.1}
\mathrm{Per}\pmb{\Lambda} \geq \frac{\lambda_\mathrm{min}^{2M}}{\prod_{i=1}^M\lambda_i}.
\end{eqnarray}

\section{Single-photon measurement probability and the permanent of Hermitian positive semidefinite matrices [proof of the equation (\ref{3}) of the main text]}

For completeness, in this section we outline the derivation of the relation between the joint single-photon measurement probability $p^{\mathrm{th}}$ at the output of a linear-optically evolved $M$-mode thermal state, and the permanent of Hermitian positive semidefinite matrices [Eq.~(\ref{3}) of the main text], following the corresponding proof of Ref.~\cite{ralph}. For this purpose we use the Husimi $Q$-function representation of the thermal state $\rho_i^\ttt{th}$~\cite{ulf, agarwal}:
\begin{eqnarray}\label{3.13}\nonumber
Q^\text{th}_i(\alpha_i)=&&\frac{1}{\pi}\avg{\alpha_i|\rho_i^\ttt{th}|\alpha_i}=\\
&&\frac{1}{\pi (\avg{n_i}+1)} \exp\left(-\frac{|\alpha_i|^2}{\avg{n_i}+1}\right),
\end{eqnarray}
which, for an $M$-mode thermal state, $\bigotimes_{i=1}^M\rho_i^\text{th}$ yields
\begin{equation}\label{3.14}
Q_{\mathrm{in}}^\text{th}(\pmb{\alpha})=\prod_{i=1}^M Q_i^\text{th}(\alpha_i),
\end{equation}
where $\pmb{\alpha}$ denotes the set of variables $\{\alpha_1, ..., \alpha_M\}$. On the other hand, a remarkable feature of the Husimi function is that for the state $\rho_{\mathrm{out}}=\mathcal{U}\rho_{\mathrm{in}}^\ttt{th}\mathcal{U}^\dagger$ it is again a product of the input functions $Q_i^\text{th}(\alpha_i)$, but of a different argument:
\begin{eqnarray}\label{3.15}\nonumber
Q_{\mathrm{out}}(\pmb{\alpha})=\frac{1}{\pi^M}\bra{\pmb{\alpha}}\rho_\mathrm{out}\ket{\pmb{\alpha}}=
\frac{1}{\pi^M}\bra{\pmb{\alpha}}\mathcal{U}\rho_\mathrm{in}^\text{th}\mathcal{U}^\dagger\ket{\pmb{\alpha}}=\\ \frac{1}{\pi^M}\bra{\pmb{\eta}}\rho_\mathrm{in}^\text{th}\ket{\pmb{\eta}}=\prod_{i=1}^M Q_i^\text{th}\left(\sum_{j=1}^M\bar{\bf{U}}_{ji}\alpha_j\right).
\end{eqnarray}
In the above equation $\ket{\pmb{\alpha}}=\bigotimes_{i=1}^M\ket{\alpha_i}$, $\ket{\pmb{\eta}}=\mathcal{U}^\dagger\ket{\pmb{\alpha}}$, and $\bar{{\bf U}}$ stands for the complex conjugate of the unitary matrix ${\bf U}$. As a result,
\begin{equation}\label{3.16}
Q_{\mathrm{out}}(\pmb{\alpha})=\frac{1}{\pi^M\prod_{i=1}^M(\avg{n_i}+1)}\exp\left(-\vec{\pmb{\alpha}}{\bf B}\vec{\pmb{\alpha}}^\dagger\right),
\end{equation}
where $\vec{\pmb{\alpha}}=(\alpha_1, ..., \alpha_M)$ stands for the row of variables $\alpha_i$, ${\bf B}={\bf U}{\pmb \zeta}{\bf U}^\dagger$, and ${\pmb \zeta}=\mathrm{diag}[1/(\avg{n_1}+1), ..., 1/(\avg{n_M}+1)]$. Meanwhile, the single-photon measurement probability $p^{\mathrm{th}}$ at the output of the linear optical circuit can be also rewritten as
\ben\label{3.18}\nonumber
p^{\mathrm{th}}=&&\text{Tr}\left[\rho_\text{out}\ket{\pmb{1}}\bra{\pmb{1}}\right]=\\
&&\pi^{M}\int_{\mathbb{C}^M}d\pmb{\alpha} Q_{\text{out}}(\pmb{\alpha})P_{\ket{\pmb{1}}\bra{\pmb{1}}}(\pmb{\alpha}).
\een
Here $\ket{\pmb{1}}=\bigotimes_{i=1}^M\ket{1}$ and $P_{\ket{\pmb{1}}\bra{\pmb{1}}}(\pmb{\alpha})$ stands for the Glauber-Sudarshan $P$ representation of the single-photon state projector $\ket{\pmb{1}}\bra{\pmb{1}}$~\cite{ulf}:
\begin{equation}\label{3.19}
P_{\ket{\pmb{1}}\bra{\pmb{1}}}(\pmb{\alpha})=\prod_{i=1}^M e^{|\alpha_i|^2}\frac{\partial^{2}}{\partial{\alpha_i}\partial{\bar{\alpha}_i}}\delta^{(2)}(\alpha_i),
\end{equation}
where $\delta^{(2)}(\alpha_i)=\delta[\ttt{Re}(\alpha_i)]\delta[\ttt{Im}(\alpha_i)]$ is the two-dimensional Dirac delta function.
Using Eq.~(\ref{3.19}), we thus rewrite Eq.~(\ref{3.18}) as
\ben\label{3.20}\nonumber
p^{\mathrm{th}}=\frac{1}{\prod_{i=1}^M(\avg{n_i}+1)}\int_{\mathbb{C}^M}d\pmb{\alpha} \exp\left(-\vec{\pmb{\alpha}} {\bf B} \vec{\pmb{\alpha}}^\dagger\right)\times \\
\prod_{i=1}^M e^{|\alpha_i|^2}\frac{\partial^{2}}{\partial{\alpha_i}\partial{\bar{\alpha}_i}}\delta^{(2)}(\alpha_i).
\een
Finally, integrating the last equation by parts we obtain the following expression:
\begin{equation}\label{3.21}
p^{\mathrm{th}}=\frac{1}{\prod_{i=1}^M(\avg{n_i}+1)}\left[\left.\prod_{i=1}^M \frac{\partial^{2}}{\partial{\alpha_i}\partial{\bar{\alpha}_i}}e^{F(\pmb{\alpha})}\right]\right|_{\alpha_i=0},
\end{equation}
where
\begin{eqnarray}\label{3.22}
F(\pmb{\alpha})=\vec{\pmb{\alpha}} {\bf D} \vec{\pmb{\alpha}}^\dagger=\sum_{i,j=1}^M {\bf D}_{ij}\alpha_i\bar{\alpha}_j,
\end{eqnarray}
with ${\bf D=I-B}$, ${\bf I}$ being the $M\times M$ identity matrix.
Note that the function $F(\pmb{\alpha})$ is a second-order polynomial in $\alpha_i$ and $\bar{\alpha}_j$, where every term is proportional to $\alpha_i\bar{\alpha}_j$, while the right-hand side of Eq.~(\ref{3.21}) corresponds to a product of derivatives of a multivariate exponential function evaluated at $\pmb{\alpha}=0$ [$e^{F(\pmb{\alpha}=0)}=1$]. In full generality, this expression is written down as a sum over products of combinations of partial derivates of the function $F(\pmb{\alpha})$ with respect to the variables $\alpha_i$ and $\bar{\alpha}_j$. Due to the quadratic from of $F(\pmb{\alpha})$ and the  evaluation at $\pmb{\alpha}=0$, we thus observe that the only terms that contribute to the final result are products of $M$ second-order derivatives of $F(\pmb{\alpha})$, where we first derive over a variable $\alpha_{i}$ followed by a derivative over $\bar{\alpha}_{j}$. Next, every $i$ and $j$ can only appear once over each product of $M$ second-order derivatives, i.e., we have instances of the form $\prod_{i=1}^M\partial^2 F(\pmb{\alpha})/\partial\alpha_{i}\partial\bar{\alpha}_{\sigma(i)}$, where $\sigma$ denotes a specific permutation of natural numbers $\{1, ..., M\}$. Because of the symmetry and the form of Eq.~(\ref{3.21}) we can see that the sum runs over all permutations $S_M$ of the set $\{1, ..., M\}$, which leads to the following expression for $p^\ttt{th}$:
\begin{equation}\label{3.23}
p^{\mathrm{th}}=\frac{1}{\prod_{i=1}^M(\avg{n_i}+1)}\sum_{\sigma \in S_M}\prod_{i=1}^M \frac{\partial^{2}F(\pmb{\alpha})}{\partial{\alpha_i\partial{\alpha_{\sigma(i)}}}}.
\end{equation}
Therefore, using Eq.~(\ref{3.22}) we find that each term in the above sum represents a product of $M$ elements of ${\bf D}$, $\prod_{i=1}^M {\bf D}_{i \sigma(i)}$.
Hence, by the definition of the permanent, we conclude that
\begin{equation}\label{3.24}
p^{\mathrm{th}}=\frac{1}{\prod_{i=1}^M(\avg{n_i}+1)}\text{Per} {\bf D}.
\end{equation}


\begin{thebibliography}{100}

\bibitem {review} P. Kok, W. J. Munro, K. Nemoto, T. C. Ralph, J. P. Dowling, G. J. Milburn, Rev. Mod. Phys. {\bf 79}, 135 (2007).

\bibitem {klm} E. Knill, R. Laflamme, G. J. Millburn, Nature {\bf 409}, 46 (2001).

\bibitem {sources} X. Ding, Y. He, Z.-C. Duan, N. Gregersen, M.-C. Chen, S. Unsleber, S. Maier, C. Schneider, M. Kamp, S. H\"{o}fling, C.-Yang Lu, J.-W. Pan, Phys. Rev. Lett. {\bf 116}, 020401 (2016).

\bibitem {circuit} J. Carolan, C. Harrold, C. Sparrow {\it et al}, Science {\bf 349}, 711 (2015).

\bibitem {boson} S. Aaronson, A. Arkhipov, Theory of Computing {\bf 9}, 143 (2013).

\bibitem {perm} S. Scheel, S. Y. Buhmann, Acta Physica Slovaca {\bf 58}, 675 (2008).

\bibitem {valiant} L. G. Valiant, Theoret. Comput. Sci. {\bf 8}, 189 (1979).

\bibitem {scott} S. Aaronson, Proc. R. Soc. A {\bf 467}, 3393 (2011).

\bibitem {bs} J. B. Spring, B. J. Metcalf, P. C. Humphreys {\it et al.}, Science {\bf 339}, 798 (2013); M. A. Broome, A. Fedrizzi, S. Rahimi-Keshari, {\it et al.}, {\it ibid.} {\bf 339}, 794 (2013); M. Tilmann, B. Daki\'{c}, R. Heilmann, S. Nolte, {\it et al.} Nat. Photonics {\bf 7}, 540 (2013); A. Crespi, R. Osellame, R. Ramponi, {\it et al.}, {\it ibid.} {\bf 7}, 545 (2013).

\bibitem {ralph} S. Rahimi-Keshari, A. P. Lund, T. C. Ralph, Phys. Rev. Lett. {\bf 114}, 060501 (2015).

\bibitem {equiv} R. J. Glauber, Phys. Rev. Lett. {\bf 10}, 84 (1963);
                 E. C. G. Sudarshan, Phys. Rev. Lett. {\bf 10}, 277 (1963).

\bibitem {scottblog} S. Aaronson, See point (4) of Comment $\#$84 at http://www.scottaaronson.com/blog/?p=2408$\#$comment-757410.

\bibitem {sch} V. S. Shchesnovich, Phys. Rev. Lett. {\bf 116}, 123601 (2016).

\bibitem {gurvits} L. Gurvits, {\it Mathematical Foundations of Computer Science}, pp. 447-458 (Springer Verlag, Berlin, 2005).

\bibitem {eigen} J. W. Demmel, {\it Applied Numerical Linear Algebra} (SIAM, Philadelphia, 1997);
                 L. N. Trefethen, D. Bau, III, {\it Numerical Linear Algebra} (SIAM, Philadelphia, 1997).

\bibitem {sample} G. E. P. Box, M. E. Muller,  Ann. Math. Stat. {\bf 29}, 610 (1958);
                  A. J. Kinderman, J. F. Monahan, ACM Trans. Mathematical Software {\bf 3}, 257 (1977);
                  C. F. F. Karney, {\it ibid} 42 1 (2016);
                  G. Marsaglia, W. W. Tsang, J. Stat. Softw. {\bf 5}, (2000).

\bibitem {chernoff} W. Hoeffding, JASA {\bf 58}, 13 (1963).

\bibitem {chernoff1} H. Chernoff, Ann. Math. Stat {\bf 23}, 493 (1952).

\bibitem {ryser} H. Ryser, {\it Combinatorial Mathematics} (Wiley, New York, 1963).

\bibitem {gurvits1} S. Aaronson, T. Hance, Quantum Information and Computation {\bf 14}, 541 (2014).

\bibitem {kaplan} W. Kaplan, {\it Advanced Calculus} (Addison-Wesley, 2002).

\bibitem {agarwal} G. S. Agarwal, {\it Quantum Opics} (Cambridge University Press, Cambridge, 2013).

\bibitem {ulf} U. Leonhardt, {\it Essential Quantum Optics} (Cambridge University Press, Cambridge, 2010).

\end{thebibliography}
\end{document}